\documentclass[12pt]{article}

\catcode`\@=11
\global\arraycolsep=2pt
\usepackage{amsbsy,amssymb,latexsym,amsfonts,amsmath}
\usepackage{graphicx,color}

\allowdisplaybreaks

\begin{document}

\begin{flushright}
\parbox{4.2cm}
{RUP-20-20}
\end{flushright}

\vspace*{0.7cm}

\begin{center}
{ \Large Conformal invariance from scale invariance in non-linear sigma models}
\vspace*{1.5cm}\\
{Yu Nakayama}
\end{center}
\vspace*{1.0cm}
\begin{center}

Department of Physics, Rikkyo University, Toshima, Tokyo 171-8501, Japan

\vspace{3.8cm}
\end{center}

\begin{abstract}
There exists a certain argument that in even dimensions, scale invariant quantum field theories are conformal invariant. We may try to extend the argument in $2n + \epsilon$ dimensions, but the naive extension has a small loophole, which indeed shows an obstruction in non-linear sigma models in $2+\epsilon$ dimensions. Even though it could have failed due to the loophole, we show that scale invariance does imply conformal invariance of non-linear sigma models in $2+\epsilon$ dimension from the seminal work by Perelman on the Ricci flow. 
\end{abstract}

\thispagestyle{empty} 

\setcounter{page}{0}

\newpage

\section{Introduction}
The advent of conformal bootstrap approaches to critical phenomena (e.g \cite{Poland:2018epd} for a review) raises a renewed interest in understanding about under which conditions the conformal symmetry emerges. Empirically, it is typically the case that scale invariance, Poincar\'e invariance (Euclidean invariance), and unitarity (reflection positivity) give rise to the enhanced conformal symmetry. Some argument supporting this empirical fact exist in even space-time dimensions, in particular two \cite{Polchinski:1987dy} and four dimensions \cite{Luty:2012ww}\cite{Dymarsky:2013pqa}\cite{Dymarsky:2014zja}\cite{Yonekura:2014tha}, but we do not have general argument in odd dimensions, say in three dimensions.\footnote{Indeed, we do have an example of scale invariant but not conformal invariant field theories such as a free $U(1)$ gauge theory in three dimensions \cite{ElShowk:2011gz}, so making the condition more precise is imperative.}

In the perturbative regime, the (non-)existence of scale invariant but not conformal field theory is closely related to the gradient nature of the renormalization group flow and the absence of the limit cycle \cite{Wallace:1974dy}\cite{Osborn:1991gm}\cite{Friedan:2009ik}\cite{Nakayama:2012nd}\cite{Fortin:2012hn}\cite{Grinstein:2013cka}\cite{Jack:2013sha}\cite{Baume:2014rla}\cite{Jack:2015tka}. Again, we do have supporting evidence for the gradient nature of the renormalization group flow in two and four dimensions. A crucial fact here is that the potential function for the gradient flow is given by Weyl anomaly coefficients at the conformal fixed point. They do exist in even dimensions but they do not exist in odd dimensions. 

Without a general argument, it may be a natural idea to explore conformal invariance in odd dimensions by using the extrapolation of $d= 2n+\epsilon$ dimensions.  Such approaches in various field theories are attempted in \cite{Polchinski:1987dy}\cite{Dorigoni:2009ra}\cite{Fortin:2011ks}. 
In this paper, we offer general discussions on how to obtain a gradient flow of the renormalization group beta function in $d=2n+\epsilon$ dimensions once we know that it is a gradient flow in $d=2n$ dimensions. This typically implies conformal invariance in (perturbative) scale invariant fixed point in $d=2n+\epsilon$ dimensions if any. 

We, however, find a small loophole in this argument, which indeed shows an obstruction in non-linear sigma models in $d = 2+\epsilon$ dimensions. The loophole is related to the question if the potential function for the gradient flow is bounded under the presence of the ambiguities in the beta functions.
Even though the simple idea could have failed due to the loophole, we can still show that scale invariance does imply conformal invariance of non-linear sigma models in $d = 2+\epsilon$ dimension from the work by Perelman on the Ricci flow \cite{Perelman:2006un}. This, on the other hand, suggests that a general argument without a loophole would be quite non-trivial: at least it should directly imply Perelman's theorem on the Ricci flow.

\section{A simple argument and possible loopholes}
We study a renormalization group flow of a  local quantum field theory with Poincar\'e invariance. The properties of the renormalization group flow is characterized by the beta functions that appear in the trace of the energy-momentum tensor.

Consider a general structure of the trace of the energy-momentum tensor (in flat space-time)
\begin{align}
T^{\mu}_{\mu} = \beta^I O_I + s_a \partial_\mu J^a + \tau_i \Box \Phi^i \ .
\end{align}
By using identities in a given field theory such as the non-conservation of the vector operator $\partial_\mu J^\mu_a = f^I_{a} O_I$, it is more convenient to rewrite the right hand side as
\begin{align}
T^{\mu}_{\mu} = B^I O_I \ . 
\end{align}
We will use this scheme to evolve the coupling constant under the renormalization group flow: $\frac{d g^I}{dt} = B^I(g)$. 
Scale invariance demands $\beta^I=0$ while conformal invariance demands $B^I=0$. If $s_a\partial_\mu J^\mu_a$ is non-zero, it is sometimes called the Virial current. In most situations, one may remove $\tau_i$ by adding local counterterms to the action, but sometimes it gives a non-trivial consequence by adding  further ambiguities in the definition of the beta functions.

In even dimensions $d=2n$, there is a general argument that at the scale invariant fixed point all $B^I$ (rather than $\beta^I$) vanish, and the conformal invariance follows. One such argument is based on the gradient property of the beta functions. It claims that the beta functions  are generated by a gradient flow:
\begin{align}
\frac{dg^I}{dt} = B^I = \chi^{IJ} \frac{\partial{a}}{\partial g^J} \ ,
\end{align}
with respect to a certain potential function $a(g)$, where we assume $\chi^{IJ}(g)$ is positive definite. If this is the case,  we can show
\begin{align}
\frac{d a}{d t} = B^I \frac{\partial a}{\partial g^I} = B^I \chi_{IJ} B^J \ge 0  \ ,
\end{align}
where $\chi_{IJ}$ is an inverse of $\chi^{IJ}$.
In other words, $a(g(t))$ is monotonically decreasing along the renormalization group flow.\footnote{Our convention is $t=\log\Lambda$ with cut-off $\Lambda$, and large $t$ corresponds to ultraviolet. Throughout the paper, we use the conventional term ``monotonically decreasing" along the renormalization group flow, but it actually means monotonically increasing with respect to $t$.} 

In $d=2n$ dimensions, $a(g)$ at the conformal fixed point is the Weyl anomaly coefficient which is positive definite. Therefore, if the theory under consideration can be deformed to be gapped, $a(g)$ cannot decrease forever. In the perturbative regime, we can argue that scale invariance demands $\frac{da}{dt}$ is (at the worst) constant, but the constant must be zero since $a(g)$ is bounded. Then the positivity of $\chi_{IJ}$ demands $B^I$ are all zero, implying that the scale invariant fixed points are actually conformal invariant.

In the literature, there have been substantial works on how to implement the above scenario in a concrete manner. We also realize that there are various subtle issues (e.g. if $\chi_{IJ}$ can remain positive definite beyond the perturbation theories). We are not going to review such issues, but we refer \cite{Nakayama:2013is} for a review.

In this paper, we simply assume the gradient flow nature of the beta functions in $d=2n$ dimensions, and we would like to see if we can extend the above analysis in $d=2n+\epsilon$ dimensions. When we use the dimensional regularization with minimal subtraction, the beta functions $\tilde{B}^I$ in $d=2n+\epsilon$ dimension and that of $d=2n$ dimensions ${B}^I$ are related by
\begin{align}
\tilde{B}^I = \epsilon k^I + B^I , 
\end{align}
where $k^I$ may depend on the operator under consideration.\footnote{For example, if we consider Yukawa-$\phi^4$ theory in $d=4-\epsilon$ dimensions, the Yukawa coupling has $k=1/2$ while the scalar quartic coupling has $k=1$.}
We also note that this simple relation only holds in a particular renormalization scheme, and we will commit ourselves to such a scheme in the following.

Let us further assume we are working in the perturbative regime so that we may regard the field space metric as a unit matrix $\chi_{IJ} = \delta_{IJ}$. Then, if $B^I$ is a gradient flow, $\tilde{B}^I$ is a gradient flow as well 
\begin{align}
\tilde{B}^I = \chi^{IJ}\frac{\partial \tilde{a}}{\partial g^J} \ ,
\end{align}
where $\tilde{a}(g) = \frac{\epsilon}{2}k^I g^I g^I + a(g)$. Note that the gradient extension might fail beyond the perturbation theory in which $\chi_{IJ}$ can be regarded as a constant.\footnote{By perturbation theory, we mean that we are close to a (conformal) fixed point. It does not necessarily mean that we are close to the Gaussian fixed point.}

Now we can repeat the analysis in $d=2n$ dimensions. If $\tilde{a}(g)$ were bounded, then we could argue $\tilde{B}^I = 0$ at the scale invariant fixed point and then we would conclude that the fixed point is conformal invariant. Here is, however, a small loophole. In $d=2n$ dimensions, $a(g)$ has a clear physical meaning such as the Weyl anomaly coefficient and it has a manifest positivity at the conformal fixed point. In $d=2n+\epsilon$ dimension, the precise physical meaning of $\tilde{a}(g)$ is unclear at this point and it could be unbounded.

Let us take a look at an example. In $d=4$ dimensions, the $\phi^4$ theory with the coupling constant $\lambda_{abcd}\phi^{a}\phi^b\phi^c\phi^d$ has the beta function
\begin{align}
B_{abcd} = \frac{1}{16\pi^2}(\lambda_{abef}\lambda_{efcd} + \lambda_{acef} \lambda_{efbd} + \lambda_{adef}\lambda_{efbc}) \  
\end{align}
so that in $d=4+\epsilon$ dimensions (being careful about the sign convention of $\epsilon$), we have
\begin{align}
\tilde{B}_{abcd} = +\epsilon \lambda_{abcd} + \frac{1}{16\pi^2}(\lambda_{abef}\lambda_{efcd} + \lambda_{acef} \lambda_{efbd} + \lambda_{adef}\lambda_{efbc}) \ . 
\end{align}
This is a gradient flow with respect to the potential
\begin{align}
\tilde{a} = +\frac{\epsilon}{2} \lambda_{abcd}\lambda_{abcd} + \frac{1}{16\pi^2} \lambda_{abcd}\lambda_{cdef}\lambda_{efab} \ . 
\end{align}
We see that $\tilde{a}$ is monotonically decreasing along the (physical) renormalization group flow. We also see that at the scale invariant fixed point, we have $\tilde{B}^a=0$ with the enhanced conformal invariance.
This is a favorable situation in which the monotonicity of $\tilde{a}$ gives proof of conformal invariance. 

%More directly, in order to obtain scale invariant but not conformal fixed point, we are asked to solve 

\section{Non-linear sigma model in $d=2+\epsilon$ dimensions}
It is widely believed that the infrared renormalization group fixed point of the scalar $\phi^4$ theory in $d=4+\epsilon$ dimensions (with negative $\epsilon$) and the ultraviolet renormalization group fixed point of non-linear sigma models in $d=2+\epsilon$ dimensions are in the same universality class if we extrapolate them to three dimensions. Since we have seen that the fixed points of $\phi^4$ theories are conformal invariant in the $d=4+\epsilon$ dimensions, we expect that the fixed points of the non-linear sigma models in $d=2+\epsilon$ dimensions are also conformal invariant.

\subsection{A direct approach}
Let us consider the non-linear sigma mode defined by the classical action
\begin{align}
S = \int d^dx G_{MN} \partial_\mu X^M \partial^\mu X^N \  
\end{align}
whose target space $\mathcal{M}$ is a $D$-dimensional compact manifold with the metric $G_{MN}(X)$. In two dimensions, it is well-known that the one-loop beta function is given by the Ricci tensor $R_{MN}(X)$ constructed out of $G_{MN}(X)$\begin{align}
B_{MN} = \frac{d G_{MN}}{d t} = R_{MN} \ 
\end{align}
up to the ambiguity of the beta functions that can be added to the right hand side (i.e. $D_M \partial_N \Phi(X) + D_N \partial_M \Phi(X)$) \cite{Friedan:1980jf}\cite{Friedan:1980jm}. This ambiguity is associated with the dilaton coupling $\mathcal{R}^{(2)}(x) \Phi(X)$ or improvement of the energy-momentum tensor. Here $\mathcal{R}^{(2)}(x)$ is the curvature of the $d$ dimensional ``world-sheet".\footnote{We would like to avoid a confusion with the target space Ricci scalar constructed out of $G_{IJ}$.}

In $2+\epsilon$ dimensions, the one-loop beta function becomes (again up to ambiguities)
\begin{align}
\tilde{B}_{MN} = -\epsilon G_{MN} + R_{MN} \ 
\end{align}
and the condition for scale invariance is
\begin{align}
\epsilon G_{MN} = R_{MN}  + D_M V_N + D_N V_M \   \label{scaleinv}
\end{align}
for a particular vector field $V^N(X)$ on $\mathcal{M}$ with the covariant derivative $D_M$. Note that the term $D_M V_N + D_N V_M$ is the diffeomorphism induced by the vector field $V_M$ (i.e. Lie derivative of the metric), so the target space is ``physically the same" with or without it.\footnote{An interesting application of this vector field can be found in \cite{Friedan:2019chh}.}

If $V_M$ is a gradient vector: $V_{M} = \partial_M F(X)$ for a certain scalar function $F(X)$ on $\mathcal{M}$, then the scale invariant fixed point is conformal invariant because one can always remove it from the above ambiguity of the beta function.
  In \cite{Polchinski:1987dy}, it was directly shown that $F=0$ when $\epsilon=0$ (even without using the ambiguity just mentioned). We would like to show a similar result when $\epsilon \neq 0$. 

Acting $D_M D_N$ on \eqref{scaleinv} and using the Bianchi identity as well as \eqref{scaleinv} again, we obtain
\begin{align}
D^M D_M R + V^M D_M R = -2R_{MN}R^{MN} + 2\epsilon R \ . \label{abc}
\end{align}
Here $R = G^{MN} R_{MN}$ is the Ricci scalar.
Let us pick a point $p$ such that $R$ takes the minimum value on $\mathcal{M}$. Since $D^M D_M R \ge 0$ and $D_M R=0$ at $p$, the left hand side of \eqref{abc} is non-negative. On the other hand, the right hand side can be rewritten as
\begin{align}
 -2R_{MN}R^{MN} + 2\epsilon R  = - 2(R_{MN}-\frac{R}{D} g_{MN})(R^{MN}-\frac{R}{D}g^{MN}) -2 R(\frac{R}{D}-\epsilon) \ .  \label{cde}
\end{align}
Here $R_{MN}-\frac{R}{D} g_{MN}$ is the traceless Ricci tensor.\footnote{The idea that the traceless Ricci tensor is useful here is inspired by the work by Hamilton \cite{Hamilton}.}
We will show that the right hand side is non-positive when $\epsilon \le 0$. 

Indeed, the trace of \eqref{scaleinv} says that $ \int d^Dx \sqrt{G} \frac{R}{D}= \epsilon \int d^Dx \sqrt{G}$, so $\epsilon$ is given by the mean curvature (divided by $D$). However, the minimum of the curvature is smaller than its mean, so $\frac{R(p)}{D} \le \epsilon \le 0 $. Thus the right hand side of \eqref{cde} is a sum of two non-positive terms, and the both must vanish.
It means that $R = D\epsilon$ is a global constant, and $R_{MN} = \frac{R}{D}g_{MN} = \epsilon g_{MN}$, showing $V_M = 0$. As we have promised $F=0$, and the scale invariant fixed point is conformal invariant. The target space is what is called the Einstein manifold.

This nice argument does not apply when $\epsilon >0$ and we cannot assign a definite sign on the right hand side of \eqref{abc}. If this were literally true, we could conclude $V_M=0$ even without considering the possibility that it could be a gradient $V_M = \partial_M F$. On the contrary, it is known that when $\epsilon >0$ there does exist a solution of \eqref{scaleinv} with non-trivial $V_M= \partial_M F$,\footnote{The first compact one was discovered by Koiso \cite{Koiso}. We will also see a non-compact example later.} and this approach must fail. We had to invent a more elaborate argument to show that scale invariance implies conformal invariance when $\epsilon > 0$. 

\subsection{A gradient approach 1}
Given success of Zamolodchikov's $c$-theorem in two dimensions \cite{Zamolodchikov:1986gt}, it is somewhat surprising that the explicit form of the monotonically decreasing $c$-function with the gradient beta functions for the non-linear sigma model was only available after the seminal work by Perelman \cite{Perelman:2006un} (see also related works \cite{Tseytlin:1987bz}\cite{Osborn:1988hd}\cite{Oliynyk:2004ey}\cite{Oliynyk:2005ak}\cite{Tseytlin:2006ak}). 

We consider the following $D$-dimensional target space action
\begin{align}
S[G,\phi] = \int d^DX \sqrt{G} e^{-2\phi} (R +  4\partial_M \phi \partial^M \phi) \  \label{effective}
\end{align}
and the $c$-function is defined by the minimum of $C[G] = -\text{inf}_\phi S[G,\phi]$ by varying $\phi$ that satisfies the normalization condition
\begin{align}
\int d^DX \sqrt{G} e^{-2\phi} = 1 \ . \label{normal} 
\end{align}
The target space action \eqref{effective} is closely related to the effective action of the string theory. There $\phi$ is identified with a string dilaton and unconstrained, but here it is important to impose the normalization condition \eqref{normal}. To make it distinguished, it is sometimes called Perelman's dilaton or minimizer $\phi_m$.

This action can be used to derive the monotonic gradient flow of the beta function
\begin{align}
G_{IM} G_{JN}\frac{e^{2\phi_m}}{\sqrt{G}} \frac{\delta C[G]}{\delta G_{IJ}} = R_{MN} + D_M \partial_N \phi_m + D_N \partial_M \phi_m \ ,  \label{Perv}
\end{align}
where Perelman's dilaton $\phi_m$ is not arbitrary but is fixed from $G_{MN}$ by minimizing $S[G,\phi]$. Remarkably this is identified with the beta function $B_{MN}$ of the metric, and in the particular scheme the renormalization group flow is generated by a gradient flow. 

Let us now argue scale invariance implies conformal invariance. In two dimensions, we see that at the scale invariant fixed point \eqref{Perv} must vanish to guarantee $\frac{dC[G]}{dt} = 0$, and it directly implies the conformal invariance. Actually, repeating the argument in the previous subsection, we can further prove $\phi_m = \text{const}$.

In $d=2+\epsilon$ dimensions, the beta function in the minimal subtraction scheme is given by
\begin{align}
B_{MN} = -\epsilon g_{MN} + R_{MN} + D_M\partial_N \Phi + D_M \partial_N \Phi \ . 
\end{align}
Here $\Phi(X)$ is an arbitrary scalar function on $\mathcal{M}$.

Now, as we discussed in section 2  we may introduce the $c$-function in $d=2+\epsilon$ dimensions by 
\begin{align}
\tilde{C}[G] = -2\epsilon \int d^Dx e^{-2\phi_m}  \sqrt{G} + C[G] \ . 
\end{align}
Here, in the first line, we do not vary $\phi$, which is already fixed in computing $C[G]$. 
This clearly gives a monotonically decreasing gradeint flow in $2+\epsilon$ dimensions 
\begin{align}
G_{IM} G_{JN}\frac{e^{2\phi_m}}{\sqrt{G}} \frac{\delta \tilde{C}[G]}{\delta G_{IJ}} = -\epsilon G_{MN}+ R_{MN} + D_M \partial_N \phi_m + D_N \partial_M \phi_m \ ,  \label{Pervv}
\end{align}
in a particular scheme where the ambiguity $\Phi$ in the beta function is fixed by Perelman's dilaton.

One may ask if this gives the proof that scale invariance implies conformal invariance in $2+\epsilon$ dimensions. The problem is that $\tilde{C}[G]$ is monotonically decreasing only for a particular $\phi_m$. We also do not know if $\tilde{C}[G]$ must be a constant at the scale invariant fixed point. 

To see the difficulty in an example, let us consider the case with $G_{MN} = \delta_{MN}$. It is somewhat surprising but crucial to notice here that $B_{MN}$ is zero only if we supplement non-trivial $\Phi= \frac{\epsilon}{4} \delta_{MN}X^M X^N$.\footnote{In mathematics literature, it is known as the Gaussian Ricci soliton.} On the other hand, when we consider the flow from \eqref{Pervv}, the Perelman's dilaton $\phi_m$ is essentially derived in two dimensions so the obvious solution here is $\phi = \text{const}$. This means that even if we have a scale invariant field theory, the $c$-function $\tilde{C}[G]$ is monotonically decreasing forever.\footnote{Indeed it is given by $-e^{\frac{-D\epsilon t}{2}}V_0$ and the would-be fixed point is a singular metric of $G_{MN} = 0$.} 
This is nothing but the loophole we have mentioned in section 2.

\subsection{A gradient approach 2}
In the seminal paper \cite{Perelman:2006un}, Perelman introduced the other monotonically decreasing functional, which he called the entropy. The direct renormalization group interpretation of Perelman's entropy in non-linear sigma models in two dimension was not obvious, but we find that it has a direct connection with conformal invariance of non-linear sigma models in $d=2+\epsilon$ dimensions. We will map the problem of finding a  scale but not conformal fixed point in the non-linear sigma model in $d=2+\epsilon$ dimension to the renormalization group flow in two-dimensions. Then we see that the stationarity of Perelman's entropy implies conformal invariance in $d=2+\epsilon$ dimensions for $\epsilon>0$.

Let us first map a scale invariant fixed point in $d=2+\epsilon$ dimensions to a non-trivial renormalization group flow in two dimensions. We will assume $\epsilon>0$. In $d=2+\epsilon$ dimensions, scale invariance implies that the metric satisfies
\begin{align}
\epsilon G_{MN} = R_{MN} + D_M V_N + D_N V_M \  \label{quasix}
\end{align}
for a certain vector field $V_N$.  Let us now define the time-dependent metric  $G_{MN}(t)$ for $t>0$ by performing time-dependent rescaling and time dependent diffeomorphism on the time-independent metric $G_{MN}$ that satisfies \eqref{quasix}: $G_{MN}(t) = \epsilon t \phi^*_V ({G}_{MN}(x))$,  where the pullback $\phi^*_V$ is induced by the diffeomorphism  $\tilde{x}^M = x^M - \epsilon^{-1}\log(t)V^M$. 

Since Ricci tensor is invariant under the rescaling of the metric (i.e. $R_{IJ}(G)= R_{IJ}(\alpha G)$), near $t=1$ the time-dependent metric $G_{MN}(t)$ satisfies the Ricci-flow equation
\begin{align}
\frac{dG_{MN}(t)}{dt} = R_{MN}(t) \ ,
\end{align} 
where $R_{MN}(t)$ is the Ricci tensor for $G_{MN}(t)$. This time evolution is nothing but the renormalization group equation in two dimensions. In this way, we have mapped a scale invariant renormalization group fixed point in non-linear sigma models $d=2+\epsilon$ dimensional to a particular renormalization group flow in two dimensions.\footnote{The discussion that follows does not explicitly use the fact that $\epsilon$ is small, but since we are neglecting the higher terms in the renormalization group beta functions, we effectively assume that $\epsilon$ is small.}

We may now want to study the renormalization group flow of $G_{MN}(t)$ in the sense of the auxiliary two-dimensional non-linear sigma model. We expect that it shows the monotonic behavior under the conventional $c$-function (or its generalization discussed in the previous section), but it is less useful in our setup because the metric typically blows up. At this point, Pelerman introduced the other monotonically decreasing quantity, which he called the entropy. Consider the functional which explicitly depends on $t$:
\begin{align}
&S[t;G_{MN}(t), \phi(t)]  \cr
&= -\int d^DX \sqrt{G(t)}\left(t(4\partial_M \phi(t) \partial^M  \phi(t) + R(t)) + 2\phi(t) -D \right)(4\pi t)^{-\frac{D}{2}} e^{-2\phi(t)} \ . \label{Ricc}
\end{align}
The claim is that this functional is monotonically decreasing along the renormalization group flow. Note that Zamolodchikov's $c$-function does not depend on $t$ explicitly so it cannot be identified with the conventional $c$-function.

We study the time-dependence of this functional under the generalized Ricci flow\footnote{The time-dependence is motivated as follows: we start with the gradient flow $ \frac{dG_{MN}}{dt} = R_{MN} + D_M \partial_N \phi + D_N \partial_M \phi$ under the fixed measure  $\sqrt{G} (4\pi t)^{-\frac{D}{2}} e^{-2\phi} $. The time-dependence of $\frac{d \phi (t)}{dt} = \frac{1}{2}\Box \phi + \frac{R}{4} -\frac{D}{4t}$ is induced from the time-independence of the measure. Then we supplement the diffeomorphism of $V_M = D_M\phi$ to make them the Ricci flow as in \eqref{Ricc}.}
\begin{align}
\frac{d G_{MN}(t)}{dt} &= R_{MN}(t) \cr
\frac{d \phi (t)}{dt} &= \frac{1}{2}\Box \phi - \partial_M \phi \partial^M \phi + \frac{R}{4} -\frac{D}{4t} \ . 
\end{align}
The direct computation gives
\begin{align}
&\frac{dS[t;G_{MN}(t), \phi(t)]}{dt} \cr
&= \int d^Dx \sqrt{G(t)} t\left(R_{MN}(t) + 2 D_M D_N \phi(t) -\frac{1}{t} G_{MN}(t)\right)^2 (4\pi t)^{-\frac{D}{2}} e^{-2\phi} \ . \label{Ricciflow}
\end{align}
Thus for $t>0$, $S[t;G_{MN}(t), \phi(t)]$ is monotonically decreasing along the renormalization group flow (i.e. monotonically increasing with respect to $t$). In particular, if $S[t;G_{MN}(t), \phi(t)]$ is stationary, it satisfies 
\begin{align}
R_{MN}(t) + 2D_M D_N f(t) -\frac{1}{t} G_{MN}(t) = 0 
\end{align}
for a particular $f$.

We emphasize here that the fixed point of $S[t;G_{MN}(t), \phi(t)]$ does not correspond to the renormalization group fixed point of two-dimensional non-linear sigma models.
Rather, it is related to a conformal invariant fixed point of non-linear sigma models in $d=2+\epsilon$ as we will explain.

Let us now argue that scale invariant fixed point in $2+\epsilon$ dimension is conformal invariant from the monotonic properties of the entropy functional. For this purpose, we maximize $S[t,G_{MN}(t),\phi(t)]$ over $\phi(t)$ under the condition $\sqrt{G} (4\pi t)^{-\frac{D}{2}} e^{-2\phi} $ is fixed. The resulting $\bar{S}[t,G_{MN}(t)] = \text{sup}_{\phi}S[t,G_{MN}(t),\phi(t)]$ is also monotonically decreasing along the renormalization group flow. Now we note that $S[t,G_{MN}(t)]$ is invariant under simultaneous scale change of $G_{MN}(t)$ and $t$ (i.e. $(G_{MN},t) \to \alpha(G_{MN},t))$. We also note that $\bar{S}[t,G_{MN}(t)]$ is invariant under the diffeomorphism on $G_{MN}(t)$ thanks to the minimization over $\phi(t)$.

Due to these two properties of $\bar{S}[t,G_{MN}(t)]$, for the Ricci-flow solution induced from the scale invariant fixed point in $d=2+\epsilon$ dimensions, we find that $\bar{S}[t,G_{MN}(t)]$ is a constant near $t=1$ since the time evolution of $G_{MN}(t)$ is generated by the scale transformation and the diffeomorphism.

On the other hand, for generic Ricci flow, we know that the time-dependence is given by \eqref{Ricciflow}. When it is stationary, it means 
\begin{align}
R_{MN}(t) + 2D_M D_N f(t) -\frac{1}{t} G_{MN}(t) = 0 , 
\end{align}
for a particular $f(t)$ that corresponds to the minimizer. However, at $t=1$ the condition can be rewritten in terms of the original metric $G_{MN}$ as 
\begin{align}
\epsilon G_{MN} = R_{MN} + 2D_M D_N F \ . \label{quasi}
\end{align}
This implies that the vector field $V_M = \partial_M F$ is a gradient and the scale invariant fixed point in $d+\epsilon$ dimension is conformal invariant.

As we have alluded above, unlike the case with $\epsilon<0$, we cannot conclude that $F$ is constant. Indeed, the manifold that satisfies the condition \eqref{quasi} is known as a gradient shrinking Ricci soliton (for positive $\epsilon$) and some non-trivial examples are available in the literature (see e.g. \cite{Koiso}). On the other hand, for negative $\epsilon$, it is known as a gradient expanding Ricci soliton, but we have already seen that they must be Einstein manifold and trivial (i.e. $F=0$).

\section{Discussions}
We have shown that scale invariance implies conformal invariance in non-linear sigma models in $d=2+\epsilon$ dimensions by using the mathematical result on the Ricci flow by Perelman. 
The monotonicity of Perelman's entropy along the renormalization group flow plays a crucial role, but it is not directly related to the renormalization group $c$-function in two dimensions because it explicitly depends on time.  It is not the renormalization group $c$-function in $d=2+\epsilon$ dimensions either because it is only defined for scale invariant theories. It, however, indicates  whether the fixed point in $d=2+\epsilon$ dimensions is conformal invariant or merely scale invariant.

It would be interesting to see if a similar function exists in other field theories than non-linear sigma models at one-loop. In particular, Perelman's idea to map the scale invariant fixed point in $d=2+\epsilon$ dimension to the non-trivial renormalization group flow in two-dimension is not conventional in physics but may be of potential significance. 

For the success of the mapping, it was crucial that the Ricci tensor is invariant under the rescaling of the metric. The similar thing may happen in one-loop gauge theories in $d=4+\epsilon$ dimensions. Suppose they are at the renormalization group fixed point 
\begin{align}
0 = -\epsilon g^{-2} + \beta_0  \ , 
\end{align}
where $\beta_0$ is a constant. We may now define the associated beta function in four dimensions from $g^{-2}(t) = \epsilon t g_*^{-2}$.
It satisfies the $d=4$ dimensional renormalization group equation
\begin{align}
\frac{d g^{-2}(t)}{dt} = \beta_0 \  
\end{align}
at one-loop. Note that it was crucial that $\beta_0$ is a constant and does not depend on $g$.

Of course, at this point, we do not know if the analog of Perelman's entropy exists. Also, we admit that the mapping may not work at the higher loop order. Both in  non-linear sigma models and gauge theories, the two-loop term (e.g. $R_{MI}R^{I}_{N}$) is not invariant under the rescaling of the coupling constant, so the naive mapping does not work. It is therefore an interesting question to show   conformal invariance of non-linear sigma models in $d=2+\epsilon$ dimensions beyond the one-loop approximation.

\section*{Acknowledgements}
This work is in part supported by JSPS KAKENHI Grant Number 17K14301.


\begin{thebibliography}{99}
%\cite{Poland:2018epd}
\bibitem{Poland:2018epd}
D.~Poland, S.~Rychkov and A.~Vichi,
%``The Conformal Bootstrap: Theory, Numerical Techniques, and Applications,''
Rev. Mod. Phys. \textbf{91}, 015002 (2019)
doi:10.1103/RevModPhys.91.015002
[arXiv:1805.04405 [hep-th]].
%187 citations counted in INSPIRE as of 05 Jun 2020


%\cite{Polchinski:1987dy}
\bibitem{Polchinski:1987dy}
J.~Polchinski,
%``Scale and Conformal Invariance in Quantum Field Theory,''
Nucl. Phys. B \textbf{303}, 226-236 (1988)
doi:10.1016/0550-3213(88)90179-4
%348 citations counted in INSPIRE as of 05 Jun 2020

%\cite{Luty:2012ww}
\bibitem{Luty:2012ww}
M.~A.~Luty, J.~Polchinski and R.~Rattazzi,
%``The $a$-theorem and the Asymptotics of 4D Quantum Field Theory,''
JHEP \textbf{01}, 152 (2013)
doi:10.1007/JHEP01(2013)152
[arXiv:1204.5221 [hep-th]].
%180 citations counted in INSPIRE as of 05 Jun 2020


%\cite{Dymarsky:2013pqa}
\bibitem{Dymarsky:2013pqa}
A.~Dymarsky, Z.~Komargodski, A.~Schwimmer and S.~Theisen,
%``On Scale and Conformal Invariance in Four Dimensions,''
JHEP \textbf{10}, 171 (2015)
doi:10.1007/JHEP10(2015)171
[arXiv:1309.2921 [hep-th]].
%94 citations counted in INSPIRE as of 05 Jun 2020


%\cite{Dymarsky:2014zja}
\bibitem{Dymarsky:2014zja}
A.~Dymarsky, K.~Farnsworth, Z.~Komargodski, M.~A.~Luty and V.~Prilepina,
%``Scale Invariance, Conformality, and Generalized Free Fields,''
JHEP \textbf{02}, 099 (2016)
doi:10.1007/JHEP02(2016)099
[arXiv:1402.6322 [hep-th]].
%49 citations counted in INSPIRE as of 05 Jun 2020


%\cite{Yonekura:2014tha}
\bibitem{Yonekura:2014tha}
K.~Yonekura,
%``Unitarity, Locality, and Scale versus Conformal Invariance in Four Dimensions,''
[arXiv:1403.4939 [hep-th]].
%7 citations counted in INSPIRE as of 05 Jun 2020

%\cite{ElShowk:2011gz}
\bibitem{ElShowk:2011gz}
S.~El-Showk, Y.~Nakayama and S.~Rychkov,
%``What Maxwell Theory in D<>4 teaches us about scale and conformal invariance,''
Nucl. Phys. B \textbf{848}, 578-593 (2011)
doi:10.1016/j.nuclphysb.2011.03.008
[arXiv:1101.5385 [hep-th]].
%91 citations counted in INSPIRE as of 05 Jun 2020



%\cite{Wallace:1974dy}
\bibitem{Wallace:1974dy}
D.~Wallace and R.~Zia,
%``Gradient Properties of the Renormalization Group Equations in Multicomponent Systems,''
Annals Phys. \textbf{92}, 142 (1975)
doi:10.1016/0003-4916(75)90267-5
%57 citations counted in INSPIRE as of 05 Jun 2020

%\cite{Osborn:1991gm}
\bibitem{Osborn:1991gm}
H.~Osborn,
%``Weyl consistency conditions and a local renormalization group equation for general renormalizable field theories,''
Nucl. Phys. B \textbf{363}, 486-526 (1991)
doi:10.1016/0550-3213(91)80030-P
%227 citations counted in INSPIRE as of 05 Jun 2020

%\cite{Friedan:2009ik}
\bibitem{Friedan:2009ik}
D.~Friedan and A.~Konechny,
%``Gradient formula for the beta-function of 2d quantum field theory,''
J. Phys. A \textbf{43}, 215401 (2010)
doi:10.1088/1751-8113/43/21/215401
[arXiv:0910.3109 [hep-th]].
%20 citations counted in INSPIRE as of 05 Jun 2020





%\cite{Nakayama:2012nd}
\bibitem{Nakayama:2012nd}
Y.~Nakayama,
%``Supercurrent, Supervirial and Superimprovement,''
Phys. Rev. D \textbf{87}, no.8, 085005 (2013)
doi:10.1103/PhysRevD.87.085005
[arXiv:1208.4726 [hep-th]].
%23 citations counted in INSPIRE as of 05 Jun 2020

%\cite{Fortin:2012hn}
\bibitem{Fortin:2012hn}
J.~F.~Fortin, B.~Grinstein and A.~Stergiou,
%``Limit Cycles and Conformal Invariance,''
JHEP \textbf{01}, 184 (2013)
doi:10.1007/JHEP01(2013)184
[arXiv:1208.3674 [hep-th]].
%91 citations counted in INSPIRE as of 05 Jun 2020

%\cite{Grinstein:2013cka}
\bibitem{Grinstein:2013cka}
B.~Grinstein, A.~Stergiou and D.~Stone,
%``Consequences of Weyl Consistency Conditions,''
JHEP \textbf{11}, 195 (2013)
doi:10.1007/JHEP11(2013)195
[arXiv:1308.1096 [hep-th]].
%32 citations counted in INSPIRE as of 05 Jun 2020


%\cite{Jack:2013sha}
\bibitem{Jack:2013sha}
I.~Jack and H.~Osborn,
%``Constraints on RG Flow for Four Dimensional Quantum Field Theories,''
Nucl. Phys. B \textbf{883}, 425-500 (2014)
doi:10.1016/j.nuclphysb.2014.03.018
[arXiv:1312.0428 [hep-th]].
%58 citations counted in INSPIRE as of 05 Jun 2020

%\cite{Baume:2014rla}
\bibitem{Baume:2014rla}
F.~Baume, B.~Keren-Zur, R.~Rattazzi and L.~Vitale,
%``The local Callan-Symanzik equation: structure and applications,''
JHEP \textbf{08}, 152 (2014)
doi:10.1007/JHEP08(2014)152
[arXiv:1401.5983 [hep-th]].
%53 citations counted in INSPIRE as of 05 Jun 2020


%\cite{Jack:2015tka}
\bibitem{Jack:2015tka}
I.~Jack, D.~Jones and C.~Poole,
%``Gradient flows in three dimensions,''
JHEP \textbf{09}, 061 (2015)
doi:10.1007/JHEP09(2015)061
[arXiv:1505.05400 [hep-th]].
%12 citations counted in INSPIRE as of 05 Jun 2020


%\cite{Dorigoni:2009ra}
\bibitem{Dorigoni:2009ra}
D.~Dorigoni and V.~S.~Rychkov,
%``Scale Invariance + Unitarity => Conformal Invariance?,''
[arXiv:0910.1087 [hep-th]].
%55 citations counted in INSPIRE as of 05 Jun 2020


%\cite{Fortin:2011ks}
\bibitem{Fortin:2011ks}
J.~F.~Fortin, B.~Grinstein and A.~Stergiou,
%``Scale without Conformal Invariance: An Example,''
Phys. Lett. B \textbf{704}, 74-80 (2011)
doi:10.1016/j.physletb.2011.08.060
[arXiv:1106.2540 [hep-th]].
%43 citations counted in INSPIRE as of 05 Jun 2020


%\cite{Perelman:2006un}
\bibitem{Perelman:2006un}
G.~Perelman,
%``The Entropy formula for the Ricci flow and its geometric applications,''
[arXiv:math/0211159 [math.DG]].
%224 citations counted in INSPIRE as of 05 Jun 2020



%\cite{Nakayama:2013is}
\bibitem{Nakayama:2013is}
Y.~Nakayama,
%``Scale invariance vs conformal invariance,''
Phys. Rept. \textbf{569}, 1-93 (2015)
doi:10.1016/j.physrep.2014.12.003
[arXiv:1302.0884 [hep-th]].
%203 citations counted in INSPIRE as of 05 Jun 2020



%\cite{Friedan:1980jf}
\bibitem{Friedan:1980jf}
D.~Friedan,
%``Nonlinear Models in Two Epsilon Dimensions,''
Phys. Rev. Lett. \textbf{45}, 1057 (1980)
doi:10.1103/PhysRevLett.45.1057
%434 citations counted in INSPIRE as of 05 Jun 2020


%\cite{Friedan:1980jm}
\bibitem{Friedan:1980jm}
D.~H.~Friedan,
%``Nonlinear Models in Two + Epsilon Dimensions,''
Annals Phys. \textbf{163}, 318 (1985)
doi:10.1016/0003-4916(85)90384-7
%507 citations counted in INSPIRE as of 05 Jun 2020


%\cite{Friedan:2019chh}
\bibitem{Friedan:2019chh}
D.~Friedan,
%``Cosmology from the two-dimensional renormalization group acting as the Ricci flow,''
[arXiv:1909.01374 [astro-ph.CO]].
%1 citations counted in INSPIRE as of 05 Jun 2020


%\cite{Hamilton}
\bibitem{Hamilton}
 R.~S.~Hamilton, ``Non-singular solutions of the Ricci flow on three manifolds". Commun. Anal. Geom. 7 (1999), 695-729.

%\cite{Koiso}
\bibitem{Koiso}
N.~Koiso, ~~On rotationally symmetric Hamilton's equation for K\"ahler-Einstein metrics, Recent topics in differential and analytic geometry," 327-337, Adv. Stud. Pure Math., 18-I, Academic
Press, Boston, MA, 1990.

%\cite{Zamolodchikov:1986gt}
\bibitem{Zamolodchikov:1986gt}
A.~Zamolodchikov,
%``Irreversibility of the Flux of the Renormalization Group in a 2D Field Theory,''
JETP Lett. \textbf{43}, 730-732 (1986)
%1384 citations counted in INSPIRE as of 05 Jun 2020


%\cite{Tseytlin:1987bz}
\bibitem{Tseytlin:1987bz}
A.~A.~Tseytlin,
%``Conditions of Weyl Invariance of Two-dimensional $\sigma$ Model From Equations of Stationarity of 'Central Charge' Action,''
Phys. Lett. B \textbf{194}, 63 (1987)
doi:10.1016/0370-2693(87)90770-2
%103 citations counted in INSPIRE as of 05 Jun 2020

%\cite{Osborn:1988hd}
\bibitem{Osborn:1988hd}
H.~Osborn,
%``String Theory Effective Actions From Bosonic $\sigma$ Models,''
Nucl. Phys. B \textbf{308}, 629-661 (1988)
doi:10.1016/0550-3213(88)90581-0
%58 citations counted in INSPIRE as of 09 Jun 2020

%\cite{Oliynyk:2004ey}
\bibitem{Oliynyk:2004ey}
T.~Oliynyk, V.~Suneeta and E.~Woolgar,
%``Irreversibility of world-sheet renormalization group flow,''
Phys. Lett. B \textbf{610}, 115-121 (2005)
doi:10.1016/j.physletb.2005.01.077
[arXiv:hep-th/0410001 [hep-th]].
%16 citations counted in INSPIRE as of 09 Jun 2020

%\cite{Oliynyk:2005ak}
\bibitem{Oliynyk:2005ak}
T.~Oliynyk, V.~Suneeta and E.~Woolgar,
%``A Gradient flow for worldsheet nonlinear sigma models,''
Nucl. Phys. B \textbf{739}, 441-458 (2006)
doi:10.1016/j.nuclphysb.2006.01.036
[arXiv:hep-th/0510239 [hep-th]].
%35 citations counted in INSPIRE as of 09 Jun 2020

%\cite{Tseytlin:2006ak}
\bibitem{Tseytlin:2006ak}
A.~A.~Tseytlin,
%``On sigma model RG flow, 'central charge' action and Perelman's entropy,''
Phys. Rev. D \textbf{75}, 064024 (2007)
doi:10.1103/PhysRevD.75.064024
[arXiv:hep-th/0612296 [hep-th]].
%32 citations counted in INSPIRE as of 05 Jun 2020

\end{thebibliography}
\end{document}